\documentclass[review]{elsarticle}

\usepackage{hyperref}
\newtheorem{thm}{Theorem}
\newproof{pf}{Proof}
\newdefinition{rmk}{Definition}

\usepackage{amsmath}
\usepackage{amssymb}
\DeclareSymbolFontAlphabet{\mathcal}   {symbols}
\usepackage{natbib}
\usepackage{graphicx}
\usepackage{algorithm}
\usepackage{algorithmic}
\usepackage{caption}
\usepackage{subcaption}
\usepackage{multirow}
\usepackage{hhline}% http://ctan.org/pkg/hhline
\usepackage[numbers]{natbib}

\DeclareMathOperator*{\argmax}{arg\,max}

\journal{Journal of Information Sciences}

\begin{document}

\begin{frontmatter}

\title{Adaptive Modularity Maximization via Edge Weighting Scheme}

%\title{Elsevier \LaTeX\ template\tnoteref{mytitlenote}}
%\tnotetext[mytitlenote]{Fully documented templates are available in the elsarticle package on \href{http://www.ctan.org/tex-archive/macros/latex/contrib/elsarticle}{CTAN}.}

%% Group authors per affiliation:
%\author{Elsevier\fnref{myfootnote}}
%\address{Radarweg 29, Amsterdam}
%\fntext[myfootnote]{Since 1880.}

%% or include affiliations in footnotes:
%\author[mymainaddress,mysecondaryaddress]{Elsevier Inc}
%\ead[url]{www.elsevier.com}

%\author[mysecondaryaddress]{Global Customer %Service\corref{mycorrespondingauthor}}
%\cortext[mycorrespondingauthor]{Corresponding author}
%\ead{support@elsevier.com}

\author[1]{Xiaoyan Lu}
\author[1]{Konstantin Kuzmin}
\author[2]{Mingming Chen}
\author[1]{Boleslaw K. Szymanski\corref{mycorrespondingauthor}}
\cortext[mycorrespondingauthor]{Corresponding author}
\ead{szymab@rpi.edu}

\address[1]{Department of Computer Science, Rensselaer Polytechnic Institute}
\address[2]{Google Inc.}

\begin{abstract}
Modularity maximization is one of the state-of-the-art methods for community detection that has gained popularity in the last decade. Yet it suffers from the resolution limit problem by preferring under certain conditions large communities over small ones. To solve this problem, we propose to expand the meaning of the edges that are currently used to indicate propensity of nodes for sharing the same community. In our approach this is the role of edges with positive weights while edges with negative weights indicate aversion for putting their end-nodes into one community. We also present a novel regression model which assigns weights to the edges of a graph according to their local topological features to enhance the accuracy of modularity maximization algorithms. We construct artificial graphs based on the parameters sampled from a given unweighted network and train the regression model on ground truth communities of these artificial graphs in a supervised fashion. The extraction of local topological edge features can be done in linear time, making this process efficient. Experimental results on real and synthetic networks show that the state-of-the-art community detection algorithms improve their performance significantly by finding communities in the weighted graphs produced by our model.
\end{abstract}


\end{frontmatter}

%\linenumbers

\section{Introduction}
Community structures are observed across a wide variety of networks, including World Wide Web, Internet, collaboration, transportation, social and biochemical networks. Many important tasks, such as data extraction, link prediction, network evolution analysis, and graph mining are based on the community structures discovered in these networks.

Modularity maximization is one of the state-of-the-art methods for community detection that has gained popularity in the last decade. It aims at discovering the partition of the network which maximizes modularity~\cite{newman2006modularity}, a widely used community quality measure proposed by Newman et al. Modularity measures the difference between the observed fraction of edges within a community and the fraction of edges expected in a random graph with the same number of nodes and the same
degree sequence. Thus, high positive modularity indicates the quality of a community structure in the network. Although modularity maximization has been widely used in many applications, in certain cases it tends to merge small communities into large ones, giving rise to the so-called \textit{resolution limit problem}~\cite{fortunato2007resolution}. In the literature, initially, it was assumed that community structure with maximum modularity is the best. Discovery of the resolution limit problem demonstrated that this is not the case. Another assumption is that the number of
communities in the given graph is unknown.

In this paper, we propose to expand the meaning of the edges that are currently used to indicate propensity of nodes for sharing the same community. In our approach this is the role for edges with positive weights while edges with negative weights indicate aversion for putting end-nodes into one community. We also propose a novel feature-based edge weighting scheme that learns how the local topological features indicate whether a given edge is intra- or inter-community using small artificial graphs similar to a network in question. Further, we demonstrate that our proposed regression model assigns weights to edges in such a way that the state-of-the-art community detection algorithms achieve higher accuracy on the produced
weighted graphs than they do on the original unweighted ones. Recent work~\cite{berry2011tolerating} shows that edge weighting scheme is capable of decreasing the upper bound on the size of communities detectable by modularity maximization. A similar approach has been adapted in~\cite{de2013enhancing} where edges are weighted according to their centrality. In contrast to~\cite{berry2011tolerating,de2013enhancing} where the edge weighting schemes are specified by experts, we develop a feature-based regression model and use labeled ground truth communities in artificial networks as training data to infer the suitable weights for edges of the input graph. These artificial networks are constructed to have degree distribution and clustering coefficient similar to the original unweighted networks. Considering the comprehensive definition of local community structures across different network instances, the regression model trained by ground truth community\footnote{If ground truth communities are not available then thanks to the small size of the artificial graph, we use communities detected algorithmically as ground truth.} in the artificial networks is therefore able to assign such weights to the edges that community detection is enhanced. Furthermore, the local topological features of edges can be extracted efficiently; so our model converts a graph into a weighted one in a time proportional to the number of edges in a network.

The experimental results on real and synthetic networks show that modularity maximization algorithms achieve higher accuracy on weighted graphs than on the original unweighted ones. For example, the optimal modularity obtained by the Fast Greedy algorithm~\cite{clauset2004finding} increases by at least 15\% on an LFR benchmark~\cite{lancichinetti2008benchmark}. We also show that our approach solves the resolution limit problem on the American college football network~\cite{evans2010clique}. In addition, the state-of-the-art community detection algorithms, including the label propagation algorithm of Raghavan et al.~\cite{raghavan2007near}, Newman's leading eigenvector method~\cite{newman2006finding}, algorithms based on random walks~\cite{pons2005computing} and the multilevel algorithm of Blondel et al.~\cite{blondel2008fast}, also improve their performance on the weighted graph produced by our approach, which validates the point that weighting graphs properly guides the algorithm to the desirable community detection results.

This paper is organized as follows. Section~\ref{sec:relatedwork} introduces the related work on modularity maximization and edge weighting schemes. Section~\ref{sec:proof} discusses the effectiveness of the edge weighting scheme. The regression model is presented in Section~\ref{sec:approach}, followed by the description of the key speedup improvements of the training algorithm. In Section~\ref{sec:experiments}, we describe the experimental results on real and synthetic networks. We close our work with conclusions presented in Section~\ref{sec:conclusions}.

\section{Related Work} \label{sec:relatedwork}
\subsection{Modularity maximization}
The goal of the modularity maximization is to discover community structure in a network by maximizing the modularity, defined as
\begin{equation} \label{eq:definition}
    Q(G, C) = \sum_{c_i \in C} \left[ \frac{|E_{c_i}^{in}| }{ |E| } - \left( \frac{d_{c_i}}{ 2 |E|}\right)^2 \right]
\end{equation}
where $G=(V,E)$ is an unweighted, undirected graph with the node set $V$ and the edge set $E$; $C=\{c_i\}$ is a partition of $G$ into communities, $c_i$ is the set of nodes in the $i$-th community, $d_{c_i}$ is the sum of degrees of nodes in $c_i$, $E_{c_i}^{in}$ denotes the set of edges residing within community $c_i$.

The modularity can be naturally extended to the networks with weighted edges by replacing the count of edges with the sum of their weights. Hence, the weighted modularity is defined as
\begin{equation} \label{eq:definition_weight}
    Q^{w}(G^w, C) = \sum_{c_i \in C} \left[ \frac{W_{c_i}^{in} }{ W } - \left( \frac{W_{c_i}}{ 2 W}\right)^2 \right]
\end{equation}
where $W$ is the sum of weights of edges in the entire graph, $W_{c_i}^{in}$ is the sum of weights of edges within community $c_i$, and the weight of a community is defined as $W_{c_i} = 2W_{c_i}^{in} + W_{c_i}^{out}$ where $W_{c_i}^{out}$ is the sum of weights of edges with exactly one endpoint inside $c_i$. The original definition of modularity is a special case of the weighted version when the weight of every edge is 1.

Many algorithms including~\cite{blondel2008fast,clauset2004finding,newman2004fast,newman2013spectral,sales2007extracting,white2005spectral} were proposed to discover communities in a network by maximizing the modularity. One interesting finding is that Newman's modularity measure is related to the broader family of spectral clustering methods ~\cite{white2005spectral}. There are two categories of spectral algorithms for maximizing modularity: one is based on the modularity matrix~\cite{newman2006finding,newman2006modularity,richardson2009spectral}, the other is based on the Laplacian matrix of a network~\cite{white2005spectral,ruan2008identifying}. The first greedy algorithm, Fast Greedy~\cite{clauset2004finding}, iteratively merges communities in the network to maximize the modularity. Initially, every node is a single community. In every step, two communities joining of which results in the largest modularity among all partitions created by temporary merging one pair of communities are merged together. After $|V|-1$ steps, there is a single community remaining in the network and there are a total of
$|V|$ partitions, each generated by a single step. Then the algorithm outputs the partition with the largest modularity. 

The greedy algorithms solve the maximization problem efficiently, yet they suffer from the resolution limit problem. This problem is defined as an increase of modularity when small well-formed (or ground truth) communities are undesirably joined together into a large community. As pointed out in \cite{fortunato2007resolution,fortunato2016community}, this problem arises because the definition of modularity does not penalize for the increase of the diameter in a community created by merging together smaller ones. In recent work \cite{chen2015new}, Chen et al. introduced a new quality metric, called {\it modularity density}, to limit such bias towards large communities. The new metric is also shown to be able to handle another known weakness of modularity, the counterproductive splitting of large communities. This is because the modularity density takes into account the density of discovered communities and penalizes the splitting of large communities. The fine-tuned $Q_{ds}$ algorithm \cite{chen2014community} was proposed to maximize this new quality metric.

\subsection{Edge weighting scheme}
Efforts have been made to improve the performance of community detection by using fine-tuned similarity measures between pairs of nodes. Such methods enhance the performance of clustering algorithms via smart edge weighting strategies. In~\cite{ruan2013efficient}, the edge weight is obtained by fusing content (pictures, tags, text) and link information (friends, followers, users) for community detection in social networks. In \cite{ma2010semi}, the authors use a set of must-link links and cannot-link links as constraints of the symmetric non-negative matrix factorization (SNMF) approach to improve the quality of discovered communities. The must-link links (edges residing within communities) and the cannot-link links (edges connecting nodes in different communities) are presumed to be known in advance. With a focus on link prediction and recommendations, Leskovec et al.~\cite{backstrom2011supervised} proposed the supervised random walk algorithm which converts an unweighted graph to a weighted graph in order to improve the performance of the random walk algorithm. Recent work~\cite{wu2015robust} proposed a random walk based approach to assign weights to nodes so that irrelevant nodes called free-riders can be excluded from some small communities. Other works including~\cite{ciglan2013community,khadivi2011network,berry2011tolerating} discuss the limitations of modularity maximization in unweighted graphs and use edge weighting schemes to improve the performance of modularity maximization algorithms.

\begin{table}[ht!]
\centering
\caption{Notations}
\begin{tabular}{ |c||l|} 
\hline 
Symbol & Meaning  \\ 
\hline \hline
$V$ & the set of nodes\\ 
\hline
$E$ & the set of edges\\ 
\hline
$d_c$ & the sum of the degrees of nodes in community $c$\\
\hline
$W$ & the sum of all edge weights\\ 
\hline
$W_{c_i, c_j}$ & the sum of weights of edges connecting communities $c_i$ and $c_j$\\
\hline
$W^{in}_c$ & the sum of weights of edges inside community $c$\\
\hline
$W^{out}_c$ & the sum of weights of edges with one endpoint in community $c$\\
\hline
$W_c$ & the weight of community c, equal to $2W^{in}_c+W^{out}_c$\\
\hline
$C$ & a partition of the graph, formed by a set of disjoint communities\\
\hline
$\Delta Q_{c_i,c_j}$ & the modulairy change caused by joining communities $c_i$ and $c_j$\\
\hline
$w_e$ & the weight of edge $e$\\ 
\hline
$x_{e}$ & the topological feature vector of edge $e$\\
\hline
$h()$ & the loss function\\
\hline
\end{tabular}
\label{tab:notations}
\end{table}

\section{Edge Weighting Scheme to Enhance Community Detection}\label{sec:proof}

As shown in~\cite{fortunato2007resolution}, for the modularity maximization to be able to find a community $c_i$ with $|E^{in}_i|$ edges inside, the following inequality must hold,
\begin{equation}
|E^{in}_i| \geq \sqrt{\frac{|E|}{2}}.
\end{equation}
where $E^{in}_i$ is the set of the edges inside community $c_i$, and $E$ is the set of edges of the entire graph. In large networks with millions of edges, the number of edges in most communities is often smaller than this lower bound. In~\cite{berry2011tolerating}, it has been shown that edge weighting scheme is capable of decreasing such theoretical bound and enhancing community detection performance in practice. Inspired by this result, we define the edge weighting scheme that \textit{enhances} particular community as follows.

\begin{rmk} An edge weighting scheme \textit{enhances} a community $c_i$ by sum of additional weights $e_i=\sum_{e\in {E^{in}_{c_i}}} (w_e - 1)>0$ if $w_e \geq 1$ for $\forall e \in E^{in}_{c_i}$,  and $w_e \leq 1 $ for $\forall e \in E^{out}_{c_i}$ while $W_{c_i} = d_{c_i}$ holds. Such a scheme is a {\it balance enhancement} if both communities connected by the cross-community edge with decreased weight are enhanced. \end{rmk}

The edge weighting scheme \textit{enhances} a community $c_i$ by increasing the weights of edges residing within this community with added total weight of $e_i>0$ , while reducing the weight of edges crossing to other communities by $2e_i$ to preserve the weight of the community $W_{c_i}$ equal to $d_{c_i}$. It is worth noting that a balanced enhancement preserves the total weight of the original graph, which is $W = |E|$. Here, we show that such balanced weighting scheme is non-decreasing modularity operation on a graph.
\begin{thm} \label{theorem:1} $Q^w(G^w, C) \geq Q(G, C)$ if the weighting scheme is balanced. \end{thm}
\begin{pf} See~\ref{appendix:1}.
\end{pf}
Although the edge weighting scheme which \textit{enhances} one community in a partition always increases this community's modularity, it does not necessarily guarantee that such enhanced partition would maximize modularity in the weighted graph. Here, we define a notion of locally maximal partition and prove that the proper edge weighting scheme can preserve such property.
\begin{rmk} Modularity of a partition $C$ is \textit{locally maximal} if the modularity decreases upon splitting any community in $C$ or joining any two communities in $C$. \end{rmk}

\begin{thm} \label{theorem:2} $\Delta Q^w_{c_i,c_j}\leq \Delta Q_{c_i,c_j}$ if $c_i,c_j$ are enhanced by the balanced edge weighting scheme. \end{thm}
\begin{pf} See~\ref{appendix:1}.
\end{pf}
\begin{thm} \label{theorem:3} If community $c$ with $d_{c} \leq \sqrt{8|E|}$, is enhanced by the balanced edge weighting scheme and split into communities $c_i,c_j$ and $\Delta Q_{c_i,c_j}\geq 0$ then also $\Delta Q^w_{c_i,c_j}\geq 0$.
\end{thm}
\begin{pf} See~\ref{appendix:1}.
\end{pf}
From Theorems~\ref{theorem:2} and~\ref{theorem:3} it follows immediately that if partition $C^*$ is locally maximal and each community $c$ satisfies
the condition $d_c \leq \sqrt{8|E|}$ and is enhanced by the balanced edge weighting scheme, the modularity of this partition is locally maximal also for the weighted graph $G^w$.

Since, by Theorem~\ref{theorem:2}, joining communities $c_i$ and $c_j$ makes change $\Delta Q^w_{c_i,c_j}\leq \Delta Q_{c_i,c_j}$, it is entirely possible that $\Delta Q^w_{c_i,c_j}\leq 0 <\Delta Q_{c_i,c_j}$. Thus, modularity maximization for the graph with the enhanced weights will avoid joining possibly well-formed communities $c_i$ and $c_j$ while the maximization on the original graph would join them. This example demonstrates that if the well-formed small communities are \textit{enhanced}, then their chances of being detected will increase. This observation motivates us to propose a regression model for assigning weights to edges so that the real ground truth communities can be \textit{enhanced} in a network.

\section{Approach} \label{sec:approach}
\subsection{Overview}
Provided a graph $G=(V,E)$, the modularity maximization problem is to find a partition of the graph that maximizes the modularity. A partition of the graph is defined as a set of disjoint communities $C = \{c_i\}$. The modularity maximization seeks to find the partition $C^*$ such that,
\begin{equation}
C^* = \argmax_{\substack{C=\{c_i\}\\\cup_i c_i = V}} \quad Q(G, C)  
\end{equation}
where $Q(G, C)$ is defined by Eq.~(\ref{eq:definition}). Since the modularity maximization problem is known to be $\mathcal{NP}$-hard problem~\cite{brandes2008modularity}, almost all proposed solutions are heuristics which do not guarantee the optimality of the partition. In this paper, we follow the same paradigm of the original modularity maximization to detect communities but seek to assign weights to the edges to improve the quality of results. To be precise, a regression model is developed to convert an unweighted graph $G$ to a weighted graph $G^{w}$ so that modularity maximization finds communities of better quality by maximizing modularity in $G^{w}$ rather than in $G$. The regression model takes the local topological features of edges as input and outputs the weight of every edge. Notations used in this paper are listed in Table \ref{tab:notations}.

\begin{figure}
    \centering    
    \includegraphics[width=\textwidth]{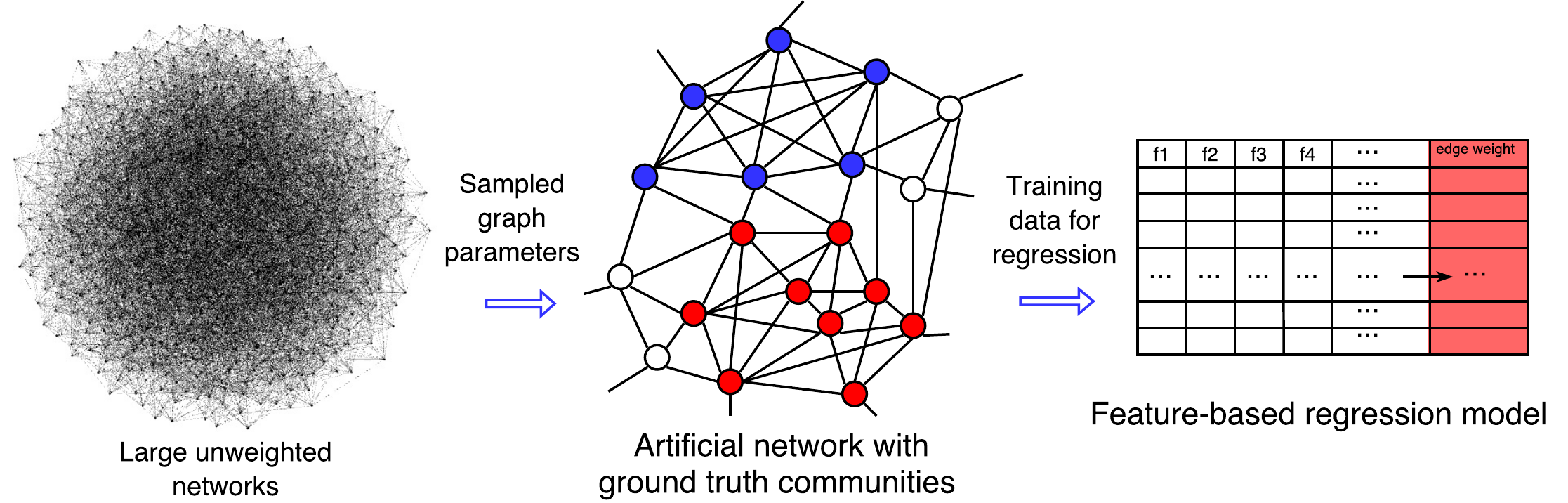}
    \caption{Illustration of the adaptive modularity maximization.}
    \label{fig:illustration}
\end{figure}

As illustrated in Figure~\ref{fig:illustration}, the proposed procedure is divided into the following three steps:
\begin{itemize}
    \item {\bf Artificial network construction} done to estimate the network parameters of the input graph and construct a similar artificial graph in which the ground truth communities are known beforehand by construction. The goal is to ensure these ground truth communities can be successfully separated in the modularity maximization process. The construction scheme is described in Section~\ref{sec:step1}.
    \item {\bf Edge feature extraction} executed on the edges of the artificial graph. The edge features are used as input to the regression model. The specific features selected by us for this purpose are discussed in Section~\ref{sec:features}.
    \item {\bf Regression on edge weights} uses a regression model to compute the edge weights such that the modularity maximization is able to separate adjacent ground truth communities in the artificial network. Section~\ref{sec:opt} covers the details of the regression model and the corresponding training algorithm. 
\end{itemize}

\subsection{Artificial network construction} \label{sec:step1}
The first step is to construct a small artificial network with the ground truth communities and with topological properties similar to the properties of the input graph. The negative edge weights are introduced to discourage the algorithm from merging ground truth communities connected by cross community edges. For clarity, we describe the usage of these ground truth communities in Section~\ref{sec:opt} and here we focus on the construction scheme.

Given a large unweighted input graph, our approach constructs an artificial network which shares degree distribution and clustering coefficients with the input graph. Specifically, multiple Stochastic Block Model (SBM) networks~\cite{holland1983stochastic} are created with high intra-block edge densities and with a few randomly chosen inter-block edges, resulting in a relatively small inter-block edge density and with blocks forming ground truth communities. Then, the edges in these SBM graphs are randomly removed from network instances until the average node degree becomes close to that of the input graph. Among all SBM network instances, the one with the average clustering coefficient closest in its value to the input graph is chosen as the final artificial network.

The ground truth communities (i.e., the nodes in blue and red in Figure~\ref{fig:illustration}) in the artificial network are used as training data to infer the parameters of the regression model. As Theorem~\ref{theorem:2} suggests, if the correct communities have been enhanced, then the probability of properly detecting these communities would increase. Therefore, the regression model incorporates these ground truth communities into the detection algorithm framework to enhance it. This approach will be discussed in detail in Section~\ref{sec:opt}.

\subsection{Extracting edge features} \label{sec:features}
Since communities are considered local structures, the second step of our approach is to extract the local topological features of every edge in the network. For each edge $e=(u,v)$ in the graph, the following local topological features are extracted efficiently from the network.
\\
\textbf{f-1}. The square root of the number of common neighbors, $\sqrt{\mathcal{N}(u) \cap \mathcal{N}(v)}$, where $\mathcal{N}(v)$ denotes the set of neighbors of node $v$. 
\\
\textbf{f-2}. The difference in clustering coefficients of the endpoints, $|c(u) - c(v)|$, where $c(v)$ denotes the clustering coefficient of node $v$. 
\\
\textbf{f-3}. Jaccard-coefficient which is defined as
\begin{equation}
\text{Jaccard}(u,v) = \frac{|\mathcal{N}(u) \cap \mathcal{N}(v) |}{|\mathcal{N}(u) \cup \mathcal{N}(v) |}
\end{equation}
\\
\textbf{f-4}. Resource allocation index which is defined as
\begin{equation}
\bigcup_{w \in \mathcal{N}(u) \cap \mathcal{N}(v)} \frac{1}{|\mathcal{N}(w)|}
\end{equation}
\\
\textbf{f-5}. Adamic-Adar index which is defined as
\begin{equation}
\bigcup_{w \in \mathcal{N}(u) \cap \mathcal{N}(v)} \frac{1}{\log|\mathcal{N}(w)|}
\end{equation}
\\
\textbf{f-6}. Relative degree ratio which is defined as
\begin{equation}
\text{rel}(u,v) = \frac{\min(|\mathcal{N}(u)|,|\mathcal{N}(v)|) }{\max(|\mathcal{N}(u)|,|\mathcal{N}(v)|)}
\end{equation}
When the degrees of nodes $u$,$v$ are equal, $\text{rel}(u,v) = 1$.

The attributes of edges or nodes, such as text content and user profiles, can also be used as features, if they are available. Using more local topological features generally leads to better accuracy because more information is embedded in these features.

\subsection{Regression model} \label{sec:opt}
As pointed out in \cite{fortunato2007resolution}, community detection algorithms based on modularity maximization tend to execute counterproductive merges of small communities into large ones. One way to handle this resolution limit problem is to cause such merging operations to decrease the modularity. 

According to the definition of weighted modularity, the change in $Q$ upon joining two communities $c_i$ and $c_j$ is
\begin{equation} \label{eq:delta_Q}
\Delta Q^w_{c_i, c_j} = \frac{W_{c_i,c_j}}{W} - \frac{W_{c_i} W_{c_j}}{2W^2}
\end{equation}
where $W_{c_i,c_j}$ is the sum of weights of the edges between $c_i$ and $c_j$, $W_{c_i} = 2W_{c_i}^{in}+W_{c_i}^{out}$ is twice the sum of weights of the edges inside community $c_i$ plus the sum of weights of the edges with exactly one edge in $c_i$, and W is the sum of weights of all edges.

To avoid the merging of some pairs of small communities $\{(c_i^1, c_i^2)\}_{i\in I}$ existing in the artificial networks
described in Section 4.2, joining them should cause a decrease of modularity, hence
\begin{equation} \label{eq:avoid}
     \Delta Q^w_{c_i^1, c_i^2} \leq 0 \hspace{2em} \text{for $i \in I$}
\end{equation}
where $I$ is the parameter defining number of pairs of small communities to be selected from the artificial network.

Using the penalty method, the optimization problem can be formulated as
\begin{align} \label{eq:opt_w}
\min_{w}  \hspace{1em} F(w) = ( \bar w - 1)^2 + \lambda_1 \sigma_w^2 + \lambda_2 \sum_{ 1 \leq i \leq I} h( \Delta Q_{c_i^1, c_i^2} )
\end{align}
where $w = \{w_e\}$ is the set of the weights of edges in the entire graph, $\sigma_w^2$ is the variance of the edge weights, $\bar w$ is the average edge weight $\frac{\sum_{e \in E} {w_e}}{|E|}$, $h(x)$ is the loss function such as the sigmoid function $h(x) = \frac{1}{1 + e^{-x}}$, $\lambda_1$ is a constant, and $\lambda_2$ is a coefficient for the penalty terms. 

The regularization term $(\bar w - 1)^2 $ ensures that the resulting average edge weight is close to 1. Using regularization on $\bar w$ directly is likely to result in very small weights that are inconvenient in community detection tasks. For the same reason, the regularization on the variance of edge weights $\sigma_w^2$ limits the total number of negative edges. When $\lambda_1 \gg \lambda_2$, the above optimization problem converges at $w_e = 1$ for $\forall e \in E$, yielding the weights of edges in an unweighted graph.

So far, we have presented an optimization method of modifying edge weights which helps avoiding improper merging of communities. However, it involves as many variables as the total number of edges and does not guarantee that edges with similar features have similar weights. Let's denote the $i$-th topological feature of one edge $e$ as $x_e^{<i>}$. The weight of an edge $e$ is obtained by the linear regression
\begin{equation} \label{eq:regression}
w_e = p_0 + \sum_{i=1}^{r} p_i x_{e}^{<i>}
\end{equation}
where $p_i$ is the parameter of the $i$-th feature, for $i=1,\ldots,r$. Let the feature vector of an edge $e$ be $x_e = (1,x_e^{<1>},x_e^{<2>},\ldots,x_e^{<r>})^T$. Then Eq.~(\ref{eq:regression}) can be rewritten as
\begin{equation} 
w_e = p^T x_{e}
\end{equation}
where vector $p=(p_0,p_1,\ldots,p_r)^T$. This way, the objective function in Eq. (\ref{eq:opt_w}) becomes a function over $p$.

The first order partial derivative of the objective function over $p_i$ is
\begin{align} \label{eq:foverp}
\frac{ \partial F(w(p))}{\partial p_i} = \frac{ \partial F(w)}{\partial w} \times \frac{ \partial w}{\partial p_i}
\end{align}
The second term on the right side of the above equation is
\begin{align} \label{eq:woverp}
\frac{ \partial w}{\partial p_i} = \left(  x^{<i>}_1,x^{<i>}_2,\ldots,x^{<i>}_{|E|} \right)
\end{align}
The first term on the right side of Eq. (\ref{eq:foverp}) is
\begin{equation} \label{eq:foverw}
\frac{ \partial F(w)}{\partial w} = \frac{\partial (\bar w - 1)^2 }{\partial w} + \lambda_1 \frac{\partial \sigma_w^2 }{\partial w} + \lambda_2 \sum_{i \in I} \frac{ \partial h(\Delta Q^w_{c_i^1, c_i^2})}{\partial \Delta Q^w_{c_i^1, c_i^2} } \frac{ \partial \Delta Q^w_{c_i^1, c_i^2}}{\partial w} 
\end{equation}
where the partial derivative $\frac{ \partial h(\Delta Q_{c_i^1, c_i^2})}{\partial \Delta Q_{c_i^1, c_i^2} }$ is obtained from the specific loss function $h()$. It is also easy to compute the partial derivative $\frac{ \partial \Delta Q^w_{c_i^1, c_i^2}}{\partial w} $ according to Eq.~(\ref{eq:delta_Q}).

\textbf{Algorithm}. To solve the optimization problem presented above, we can apply a quasi-Newton method, such as the Broyden-Fletcher-Goldfarb-Shanno (BFGS) algorithm~\cite{nocedal2006numerical}, which requires only the first derivative of the objective function to find the optimal result. The pseudo code of the training algorithm is presented in Algorithm \ref{algo:1}.

During the training phase, $|I|$ pairs of ground truth communities $\{c_i^1,c_i^2\}_{i \in I}$ can be chosen randomly from the artificial network assuming that the ground truth communities are provided. One efficient way to obtain the required number of pairs of ground truth communities is to sample adjacent communities in the artificial network randomly until $|I|$ pairs are collected. As indicated by Theorem~\ref{theorem:3}, small communities are preferred to large communities here. Hence, we can set an upper bound on the size of the chosen communities. After the parameters are inferred, the regression model which assigns weights to edges can be applied to enhance the performance of community detection algorithms.

\begin{algorithm} 
\caption{Regression Model Training Algorithm}
\begin{algorithmic}[1] \label{algo:1}
\STATE Initialize p
\FOR {each edge $e$}
\STATE $x_e \leftarrow$ extracted features of edge $e$
\STATE $w_e \leftarrow p^T x_e$
\ENDFOR
\STATE $tol \leftarrow 0.0001$
\STATE Construct $\{c_i^1,c_i^2\}_{i\in I}$
\REPEAT
\STATE Compute $\frac{ \partial F}{\partial w}$ using Eq. (\ref{eq:foverw})
\STATE $ \frac{ \partial F}{\partial p} \leftarrow \frac{ \partial F}{\partial w} \times \frac{ \partial w}{\partial p} $
\STATE Update $p$ via one BFGS step
\FOR {each edge $e$}
\STATE $w_e \leftarrow p^T x_e$
\ENDFOR
\UNTIL {$\|\frac{ \partial F(w(p))}{\partial p}\| < tol$ or the maximum number of iterations is made}
\end{algorithmic}
\end{algorithm}

The \textbf{time complexity} of Algorithm 1 is $O(k(|I|+|E_a|))$ where $k$ is the number of BFGS iterations before the algorithm converges, $|I|$ is the number of constraints in Eq. (\ref{eq:avoid}) and $|E_a|$ is the total number of edges in the artificial graph. In order to accelerate the computation, we adopt the following key speedup improvements.

To compute the change of modularity upon joining two communities, the weights of all edges in the artificial graph need to be summed up which takes significant amount of time in each BFGS step. The summation of weights is
\begin{equation} \label{eq:sumofweight}
    W = \sum_{e\in E_a} {w_e} = \sum_{e\in E_a} p^T x_e = p^T \sum_{e\in E_a} {x_e}
\end{equation}
which can be calculated efficiently because $\sum_e {x_e}$ needs to be computed only once at the outset of the optimization process, and $W$ is re-computed as the inner product of $p$  and $\sum_e {x_e}$ in every iteration. The sums of weights of the edges related to each community $c$, such as $W^{in}_c$ and $W^{out}_c$, and the variance of edge weights $\sigma_w^2$ can also be computed in the similar manner. Note that such speedup can be achieved because we intentionally use linear regression to compute the edge weights in Eq. (\ref{eq:regression}). Otherwise, if non-linear regression function is used to obtain the edge weights, Eq. (\ref{eq:sumofweight}) does not hold and it generally takes more time to
obtain the sums of weights.

In our algorithm, the edges with both endpoints not in any communities in pairs $\{c_i^1, c_i^2\}$ for $i\in I$ are not involved in the computation of every BFGS iteration. The number of edges involved in every BFGS iteration is at most $2|I|Z$ where $Z$ is the average number of edges in communities in pairs $\{c_i^1, c_i^2\}$. So, the time complexity is reduced to $O(|E_a| + k|I| \times 2|I|Z) = O(|E_a| + k|I|^2Z)$. In practice, this accelerated algorithm provided at least a 50-fold speedup compared to Algorithm 1. 

\textbf{Interpretation of edge weight in social networks}. Edges are usually considered equally important in many community detection applications. Then, would not be the relationships between individuals also equally important in respect to forming communities in social networks? In real-world cases, one may know a lot of people, meet with them regularly, but trust only a few. The weight of a connection could be interpreted as the strength of the trust between people, or the strength of their social influence on each other. Social influence inferring has been studied in \cite{goyal2010learning,tang2009social}. Compared to these publications, our work focuses on assigning the edge weights in a way to assist in the formation of communities rather than to explain the spreading of opinions or ideas by social influences. Compared to other edge weighting schemes~\cite{ciglan2013community,khadivi2011network,berry2011tolerating}, the proposed regression model learns the edge weighting scheme from real ground truth communities in a supervised fashion. In addition, our work assigns one-dimensional weight to edges as a scalar quantitative measure, yet the weight could be extended to be a multi-dimensional measure of the strength of influence or trust in different contexts.

It is worth noting that our approach is a novel pre-processing tool to enhance community detection algorithms in most cases, even if they are not based on modularity maximization. However, since the proposed edge weighting scheme aims at improving the modularity maximization approaches, community detection algorithms based on other principles are not guaranteed to perform better on the weighted networks than they do on the original unweighted networks.

\section{Experimental Results}
\label{sec:experiments}
In this section, we describe the experimental results obtained on real and synthetic networks. We compare the accuracy of modularity maximization algorithms running on original unweighted graphs and weighted graphs produced by our model. The experimental settings and evaluation metrics are explained in Section \ref{sec:conf}. The experimental results on synthetic and real networks are presented in Section \ref{sec:syn_real}.

\subsection{Simulation configurations}
\label{sec:conf}
To evaluate the performance, the state-of-art greedy modularity maximization algorithm, Fast Greedy~\cite{clauset2004finding}, is executed on several real and synthetic networks. The regression model is trained by sampling the ground truth communities in artificial Stochastic Block Model (SBM) networks~\cite{holland1983stochastic} in which the ground truth communities are complete, dense and well-defined~\cite{peel2016ground}. In the SBM networks, nodes are connected to one another with particular edge densities, depending on their membership in the pre-defined communities. The artificial SBM network used by our model is constructed as follows: multiple SBM network instances are created with a high intra-edges density and a random, relatively small inter-edge density. Then, edges are randomly removed from network instances until the average node degree becomes close to that of the input graph. Among all SBM instances, the one with the average clustering coefficient closest in its value to the value of this coefficient in the input graph is used to train the regression model. The convergence of our training algorithm and the construction of SBM networks take only a few seconds.

The details of the tested networks are summarized in Table~\ref{table:summary}. The regression model converts each graph into a weighted one. Then, the Fast Greedy algorithm~\cite{clauset2004finding} is executed to detect communities in both the weighted and unweighted networks. We compare the detected communities with the given ground truth communities to compute the quality measures. The ground truth communities in real networks are often determined by the specific label of nodes. Although the goal of the community detection differs from the discovery of meta-data of nodes~\cite{peel2016ground}, we consider such labels to be a strong sign of the existence of some valid partitions.

Let the ground truth partition of the graph be denoted as $GN=\{g_1, g_2,\ldots\}$ where $g_i$ is a single ground truth community. The following evaluation metrics measure the similarity between the produced partition $C$ and ground truth partition $GN$.\\
\textbf{Variation of Information} (VI)~\cite{good2010performance} measures the similarity between $C$ and $GN$ based on information theory
\begin{equation}
VI(C, GN) = H(C) + H(GN) - 2I(C, GN)
\end{equation}
where $I(C,GN)=H(C) + H(GN) - H(C,GN)$ is the \textit{Mutual Information}, and $H()$ is the entropy function defined as
\begin{align}
H(C) &= -\sum_{c_i \in C} p(c_i) \log{p(c_i)} = -\sum_{c_i \in C} \frac{|c_i|}{|V|} \log{ \frac{|c_i|}{|V|} }\\
H(C,GN) &= -\sum_{c_i \in C, g_i \in {GN}} p(c_i,g_i) \log{p(c_i,g_i)} \nonumber\\
&= -\sum_{c_i \in C, g_i \in {GN}} \frac{|c_i \cap g_i|}{|V|} \log{ \frac{|c_i \cap g_i|}{|V|} }
\end{align}
\textbf{Normalized Mutual Information} (NMI)~\cite{wagner2007comparing} is defined as
\begin{equation}
NMI(C,GN)= \frac{2I(C,GN)}{H(C) + H(GN)}
\end{equation}
\textbf{F-measure}~\cite{wagner2007comparing} is given as
\begin{equation}
\textit{F-measure}(C,GN) = \frac{1}{|V|}\sum_{c_i \in C} |c_i| \max_{g_i \in GN} \frac{2|c_i \cap g_i|}{|c_i|+|g_i|}
\end{equation}
\textbf{Adjusted Rand Index} (ARI)~\cite{hubert1985comparing} computes the similarity by comparing all pairs of nodes that are assigned to the same or different communities in $C$ and $GN$. Specifically, ARI is defined as
\begin{equation}
ARI(C,GN) = \frac{  \sum_{ij}{\binom{|c_i \cap g_j|}{2}} - \frac{ [\sum_i {\binom{|c_i|}{2}} \sum_j{ \binom{|g_j|}{2} }   ] } { \binom{|V|}{2} } } {  \frac{1}{2} [\sum_i \binom{|c_i|}{2} + \sum_j \binom{|g_j|}{2} ]  - \frac{ [\sum_i {\binom{|c_i|}{2}} \sum_j{ \binom{|g_j|}{2} }   ] } { \binom{|V|}{2} } }.
\end{equation}
\textbf{Modularity Density}~\cite{chen2014community} is a measure of the quality of communities in a network. Like the original modularity, it does not need the ground truth. The formal definition is
\begin{align}
Q_{ds} &= \sum_{c_i \in C} \left[ \frac{|E_{c_i}^{in}|}{|E|} d_{c_i} - \left( \frac{2|E_{c_i}^{in}| + |E_{c_i}^{out}|}{2|E|} d_{c_i} \right)^2 - \sum_{\substack{c_j \in C\\c_j\neq c_i}} \frac{|E_{c_i,c_j}|}{2|E|} d_{c_i,c_j}
\right]\\
d_{c_i} &= \frac{2|E_{c_i}^{in}|}{|c_i|(|c_i|-1)} \quad \quad \quad \quad d_{c_i,c_j} = \frac{|E_{c_i,c_j}|}{|c_i||c_j|} 
\end{align}
where $d_{c_i}$ is the internal density of community $c_i$, and $d_{c_i,c_j}$ is the pair-wise density between community $c_i$ and community $c_j$.

In addition, we evaluate the execution time of the training of the regression model and the additional time needed to convert an unweighted graph into a weighted one. We does not report the time cost of community detection, which depends on the specific algorithms. Hence, the reported time cost consists of two parts: (i) Training: the time cost to infer all the parameters of the regression model from the artificial graph; (ii) Weighting: the time cost to compute the weights of every edge in the original graph. Note that both parts include the I/O cost of loading the network files from disk and the edge topological feature extraction time.

\begin{table}[!t]
\caption{Summary of the networks}
\label{table:summary}
\centering
\setlength\tabcolsep{2pt}
\begin{tabular}{ c|c|c|c|c|c }
 \hline
No. & Network & \#Nodes & \#Edges & Type & Ref.\\
 \hline
1 & American college football & 115 & 613 & Real & ~\cite{evans2010clique}\\
2 & LFR benchmark & 5000 & $\approx$35000  & Synthetic & ~\cite{lancichinetti2008benchmark}\\
3 & Amazon product co-purchasing network & 334863 & 925872 & Real & ~\cite{yang2015defining}\\
4 & DBLP collaboration network & 317080 & 1049866 & Real & ~\cite{yang2015defining}\\
 \hline
\end{tabular}
\end{table}

\subsection{Performance on synthetic and real networks}
\label{sec:syn_real}

\subsubsection{LFR benchmark}
The LFR benchmark networks~\cite{lancichinetti2008benchmark} serve as one of the standards for the evaluation of community detection algorithms. The properties of the network generated from the benchmark are defined by the following three parameters: $\gamma$ which is an exponent of the node degree in the power law distribution, $\beta$ which is an exponent of the community size in the power law distribution, and $\mu$ which is the mixing parameter that defines the fraction of edges originating in a community that have one endpoint outside of it. In our experiments, every LFR benchmark network has 5,000 nodes with the average node degree 15 and the maximum node degree 50. The exponents $\gamma$ and $\beta$ are set as 2 and 1 respectively and the mixing parameter $\mu$ takes two values, 0.45 and 0.5, which are quite challenging because high values of the mixing parameter are likely to result in loose community structures. Considering the randomness in the generation of synthetic networks, 10 network instances are constructed for each $\mu$ value.

We evaluate the modularity maximization performance on the original unweighted LFR benchmark networks. As seen in Table~\ref{table:LFR}, the performance of the Fast Greedy (FG) algorithm~\cite{clauset2004finding} has been significantly improved by maximizing the modularity on the weighted networks instead of on the original unweighted graph. The F-measure is improved by nearly 40\% and the NMI metric is improved by 25\% in all cases. In Table~\ref{table:LFR}, the modularity $Q$ and modularity density $Q_{ds}$ values are all computed over the original unweighted LFR benchmark networks. This surprising result shows that the execution of Fast Greedy algorithm on weighted graph can improves the $Q_{ds}$ value for the corresponding unweighted graph. In other words, the weighted edges allows the greedy algorithm to escape from local maximum of $Q_{ds}$ and get better value of it on the original unweighted graph. Also, using the edge weighting scheme, the Fast Greedy algorithm which maximizes the weighted modularity performs better than the fine-tuned $Q_{ds}$ algorithm~\cite{chen2014community}. In addition, we compared our approach to the previously published algorithm CNM~\cite{berry2011tolerating}. The introduced here edge weighting scheme achieves an average 85\% Jaccard-index score while the CNM algorithm obtains the average score of 82\% on 10 different realizations of the LFR benchmark networks using the same parameters, with the mixing parameter set to $\mu=0.5$. The remaining specific construction parameters of these LFR benchmark networks and the definition of Jaccard-index can be found in~\cite{berry2011tolerating}.

\begin{table}[!t]
\caption{Metric values characterizing the community structures computed over the original unweighted LFR benchmark networks but discovered by different algorithms. FG: Fast Greedy modularity maximization algorithm on the original unweighted graphs. FG-w: Fast Greedy modularity maximization algorithm running on the weighted graphs produced by our model.}
\label{table:LFR}
\centering
\setlength\tabcolsep{2pt}
\begin{tabular}{|c|c||c|c|c|c|c|c|}
\hline
$\mu$ & Method & VI & NMI & F-measure & ARI & $Q$ & $Q_{ds}$ \\
\hline
\hline
\multirow{3}{*}{0.45} & FG &3.2135&0.5953&0.3379&0.2355&0.4214&0.0366\\
\hhline{~-------}
& Fine-Tuned $Q_{ds}$ & 1.1523 & 0.8925 & 0.8806 & 0.7337 & 0.4536 & 0.1632\\
\hhline{~-------}
& FG-w & 0.0137 & 0.9987 & 0.9990 & 0.9972 & 0.5152 & 0.1668\\
\hline
\multirow{3}{*}{0.5} & FG &3.5187&0.5481&0.2937&0.1993&0.3739&0.0274\\
\hhline{~-------}
& Fine-Tuned $Q_{ds}$ & 1.9677 & 0.8036 & 0.7489 & 0.4984 & 0.3563 & 0.1196\\
\hhline{~-------}
& FG-w & 0.0678 & 0.9934 & 0.9950 & 0.9864 & 0.4625 & 0.1381\\
\hline
\end{tabular}
\end{table}
\begin{figure}
    \centering
    \begin{subfigure}[b]{0.3\textwidth}
        \centering
        \includegraphics[width=\textwidth]{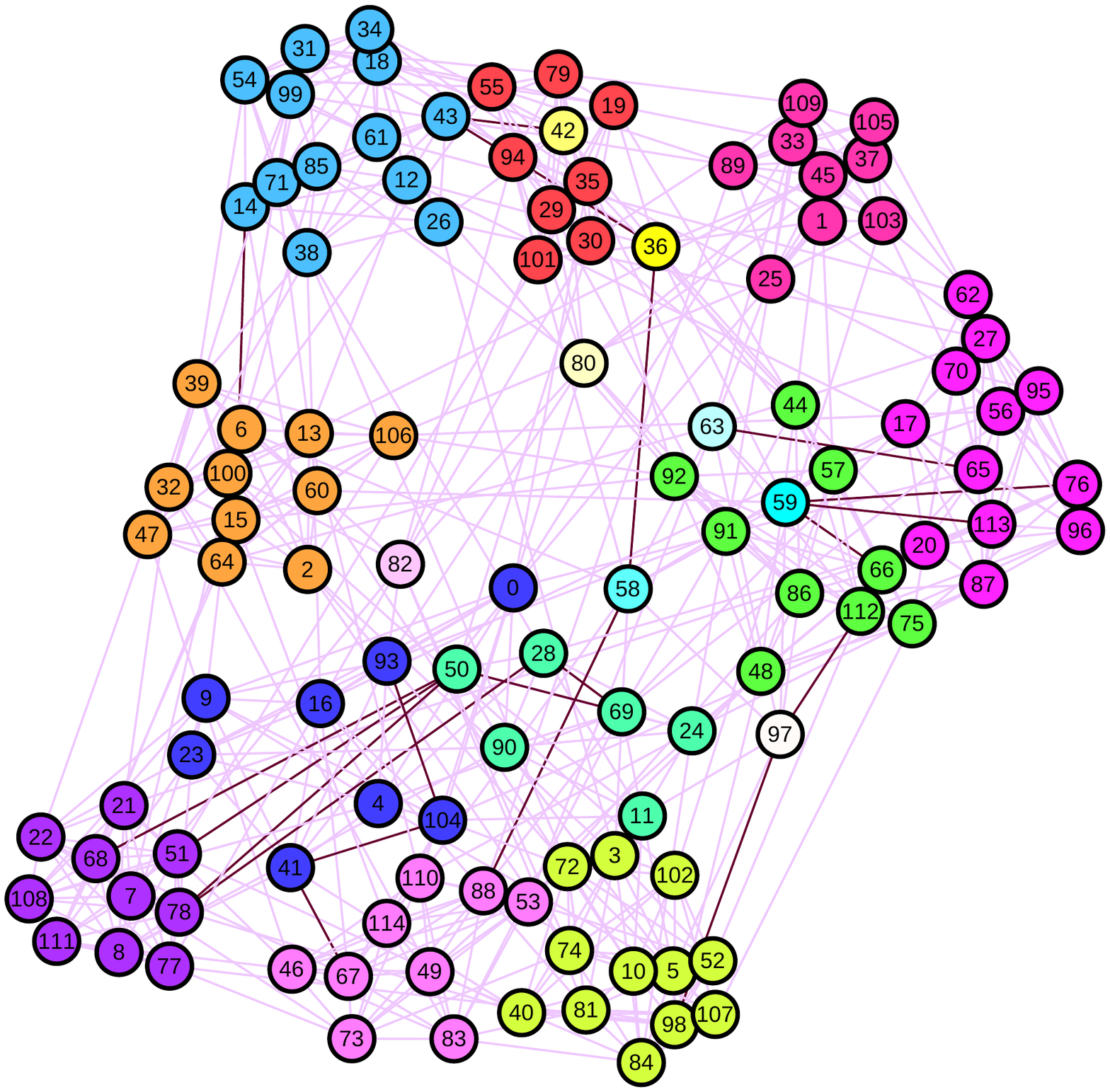}
        \caption{}
        \label{fig:football_groundtruth}
    \end{subfigure}
    \begin{subfigure}[b]{0.3\textwidth}
        \centering
        \includegraphics[width=\textwidth]{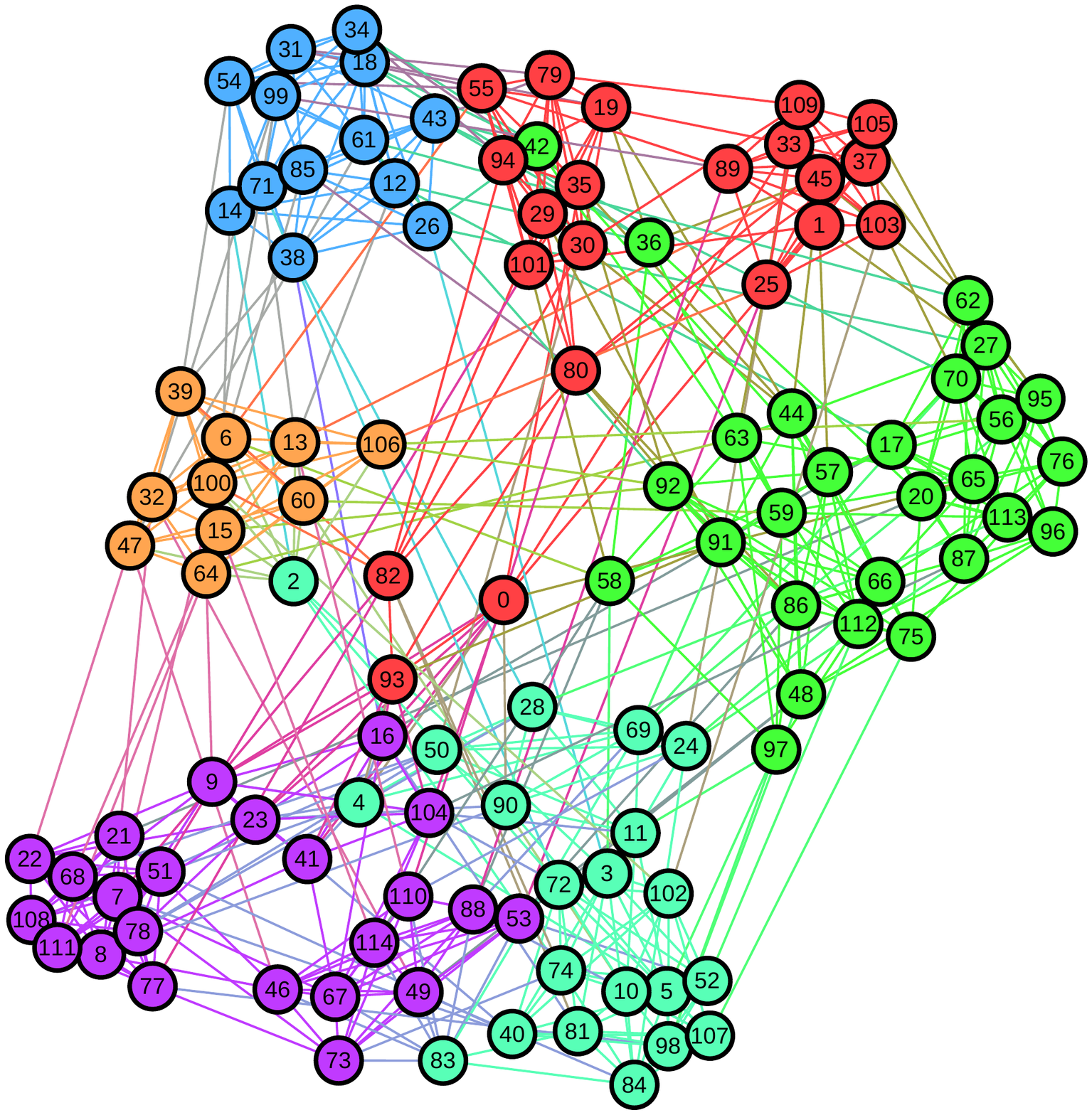}
        \caption{}
        \label{fig:football_orig}
    \end{subfigure}
    \begin{subfigure}[b]{0.3\textwidth}
        \centering
        \includegraphics[width=\textwidth]{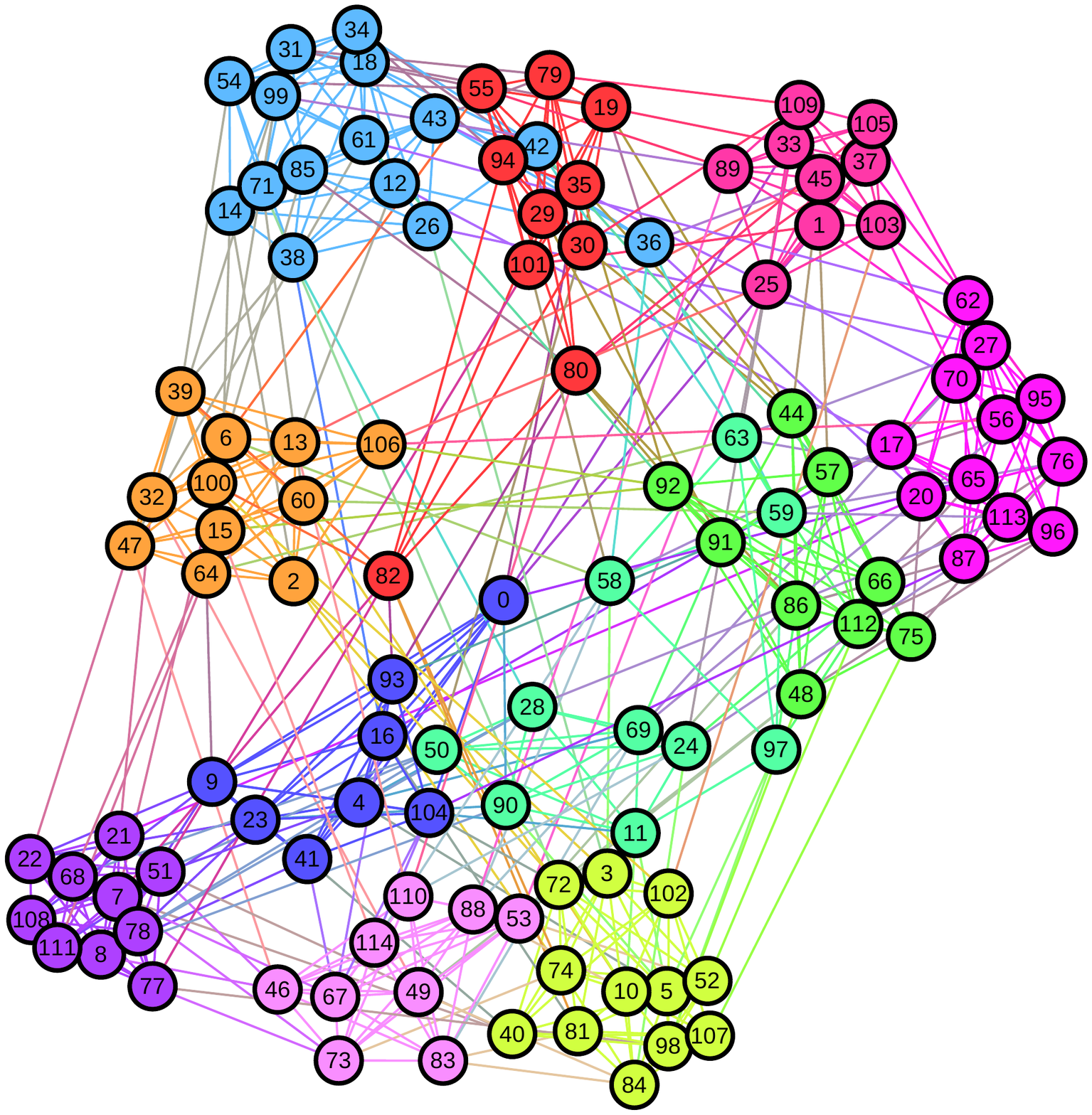}
        \caption{}
        \label{fig:football_better}
    \end{subfigure}
    \caption{The communities detected by the Fast Greedy algorithm~\cite{clauset2004finding} in the American college football network~\cite{evans2010clique}. Nodes are colored according to communities to which they have been assigned. (a) 19 ground truth communities defined as 11 conferences and 8 independent teams. Edges in black are assigned negative weights by the edge weighting scheme. (b) 6 communities detected on the unweighted graph by the modularity maximization method. (c) 11 communities detected on the weighted graph by the modularity maximization method.}
    \label{fig:football_convergence}
\end{figure}

\subsubsection{American college football network} \label{sec:football}
The American college football network~\cite{evans2010clique} consists of 115 nodes representing college football teams playing in a league with 11 conferences. Two teams are linked if they have played with each other in the year 2000 season. The teams in each of the 11 conferences can be treated as one community because they play with each other often. There are 8 independent teams (not members of any conference), each forming a single community. 19 ground truth communities are shown in Figure~\ref{fig:football_convergence}.a with each color representing a single community. However, only 6 communities are detected by the Fast Greedy algorithm on an unweighted graph as shown in Figure~\ref{fig:football_convergence}.b because some adjacent ground truth communities are joined together.

\begin{table}[!t]
\caption{Metric values characterizing the community structures computed over the original unweighted American college football network but discovered in either the original unweighted graph or the corresponding weighted graph produced by our model. FG: Fast Greedy algorithm~\cite{clauset2004finding}, LE: leading eigenvector method~\cite{newman2006finding}, LP: label propagation algorithm~\cite{raghavan2007near}, RW: community detection based on random walks~\cite{pons2005computing}, ML: multilevel algorithm~\cite{blondel2008fast}, NMI: normalized mutual information, ARI: adjusted rand index.}
\label{table:football}
\centering
\setlength\tabcolsep{2pt}
\begin{tabular}{ |c|c|c|c|c|c|c| }
 \hline
Metric & Graph & FG & LE & LP & RW & ML \\
 \hline
 \multirow{2}{*}{NMI} & Original &0.58528&0.58140&0.76962&0.83833&0.83391\\
 & Weighted & 0.91117&0.85903&0.92635&0.91117&0.87272\\
\hline
 \multirow{2}{*}{ARI} & Original &0.49333&0.49441&0.71749&0.86938&0.85815\\
 & Weighted &0.94723&0.88982&0.91539&0.94723&0.90085\\
 \hline
 \multirow{2}{*}{$Q$} & Original &
 0.56860&0.49326&0.57668&0.60337&0.60503\\
 & Weighted & 0.60140&0.59338&0.57315&0.60140&0.60356\\
 \hline
 \multirow{2}{*}{$Q_{ds}$} & Original &0.15877&0.13661&0.21106&0.23650&0.23626\\
 & Weighted &0.25696&0.23893&0.24025&0.25696&0.24889\\
 \hline
\end{tabular}
\end{table}

The regression model converts the original unweighted graph to a weighted graph where the edges with negative weights are marked in black in Figure~\ref{fig:football_convergence}.a. On the weighted graph, the Fast Greedy algorithm can find 11 league communities, each containing one individual conference, although it allocates the independent teams to some of these league communities. The regression model is trained by sampling the ground truth communities on the artificial SBM network, which is constructed to be similar to the Football network. The training process takes approximately 10 seconds on a machine with a single 2.5GHz CPU.

As illustrated in Table~\ref{table:football}, in addition to the Fast Greedy modularity maximization algorithm, the state-of-the-art community detection algorithms, including label propagation algorithm by Raghavan et al.~\cite{raghavan2007near}, Newman's leading eigenvector method~\cite{newman2006finding}, the algorithm based on random walks~\cite{pons2005computing} and the multilevel algorithm by Blondel et al.~\cite{blondel2008fast}, also demonstrate improved performance on weighted graphs produced by our method\footnote{In the experiments, the edges with negative weight are removed from the graph because some community detection algorithms are not able to handle negative weights due to the algorithm design or implementation.}. This result additionally supports our claim that properly weighting a graph can lead to an improved quality of community detection.

For a fair comparison, regardless of whether the partition of the graph is determined with or without the edge weights, both the modularity $Q$ and modularity density $Q_{ds}$ are computed over the unweighted graph, i.e., edge weights are all set to 1. Hence, a better $Q$ or $Q_{ds}$ found on the weighted graph indicates the edge weights allow the maximization algorithm to avoid the inferior local optima. The NMI and ARI measures indicate that the communities detected in the weighted graph are generally accurate. However, from the aspect of modularity, for three algorithm, LP, RW and ML, such communities may be evaluated as inferior (as they have slightly lower modularity) than the communities discovered in the original unweighted graph. Consequently, even if the maximum modularity is reached in the original unweighted graph, the resulting communities are still not likely to match the ground truth. In contrast, the modularity density $Q_{ds}$\footnote{Note that the modularity density values are all computed over the original Football network.} of the communities detected in the weighted one is higher than in the original unweighted graphs, which means that it accurately measures the quality of these communities. The proposed edge weighting scheme leads to a higher modularity density $Q_{ds}$ in all cases because the weighted edges allows the greedy algorithm to escape from local maximum of $Q_{ds}$ and get better value of it on the original unweighted graph. 

\subsubsection{Large Networks}
\begin{figure}
    \centering
    \begin{subfigure}[b]{0.45\textwidth}
        \centering
        \includegraphics[width=\textwidth]{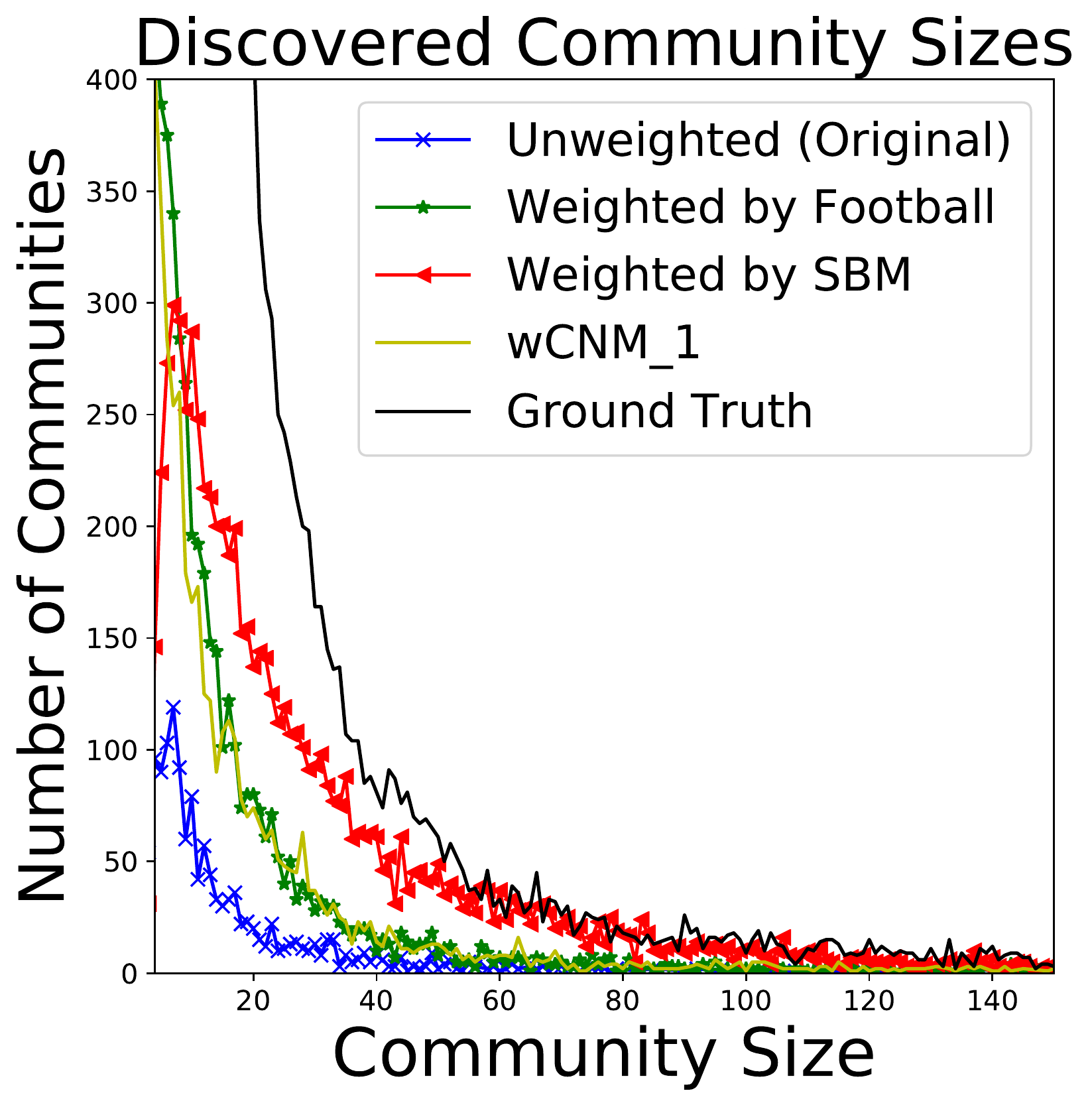}
        \caption{Amazon}
        \label{fig:1}
    \end{subfigure}
    \begin{subfigure}[b]{0.45\textwidth}
        \centering
        \includegraphics[width=\textwidth]{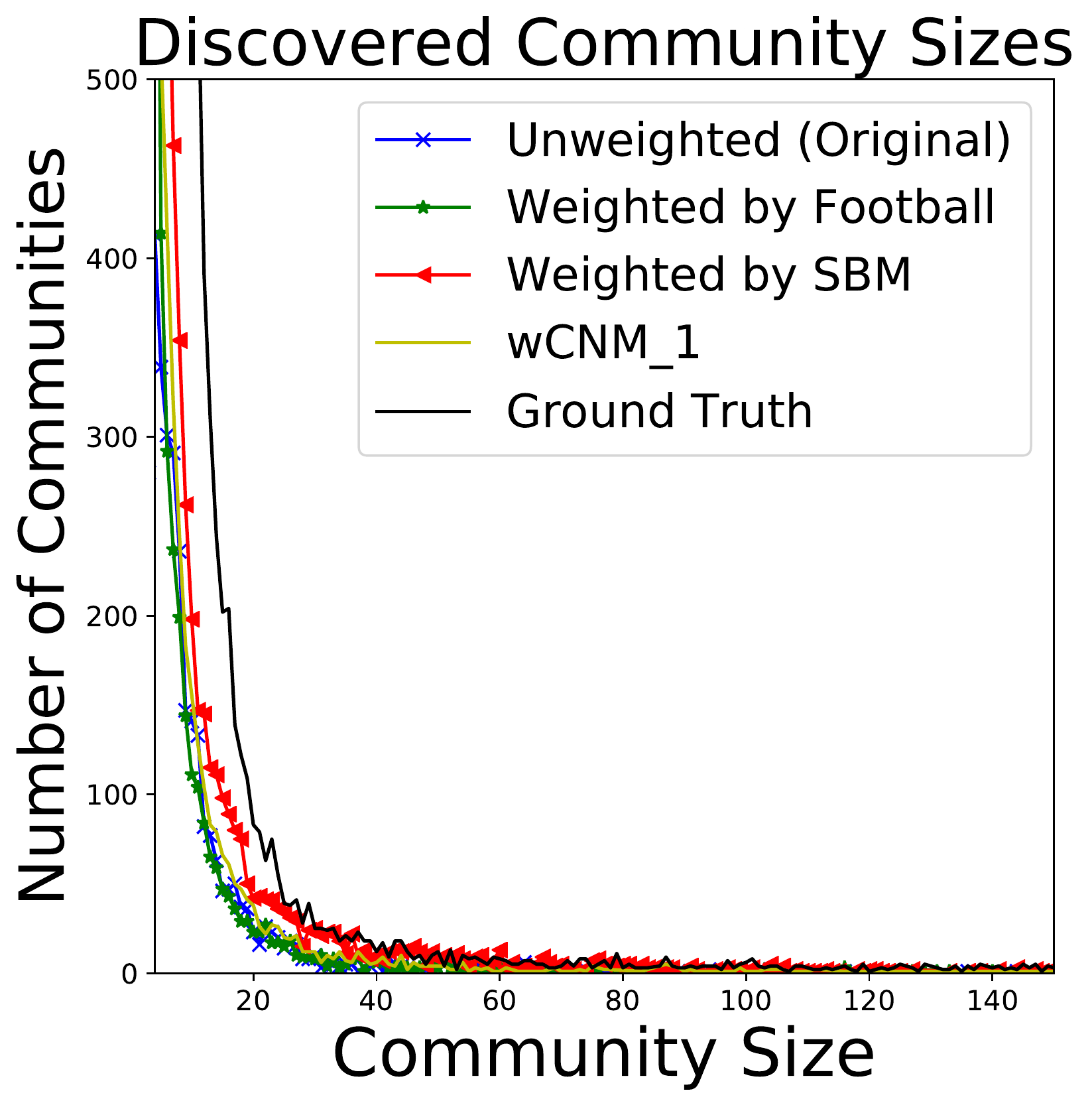}
        \caption{DBLP}
        \label{fig:2}
    \end{subfigure}
    \begin{subfigure}[b]{0.51\textwidth}
        \centering
        \includegraphics[width=\textwidth]{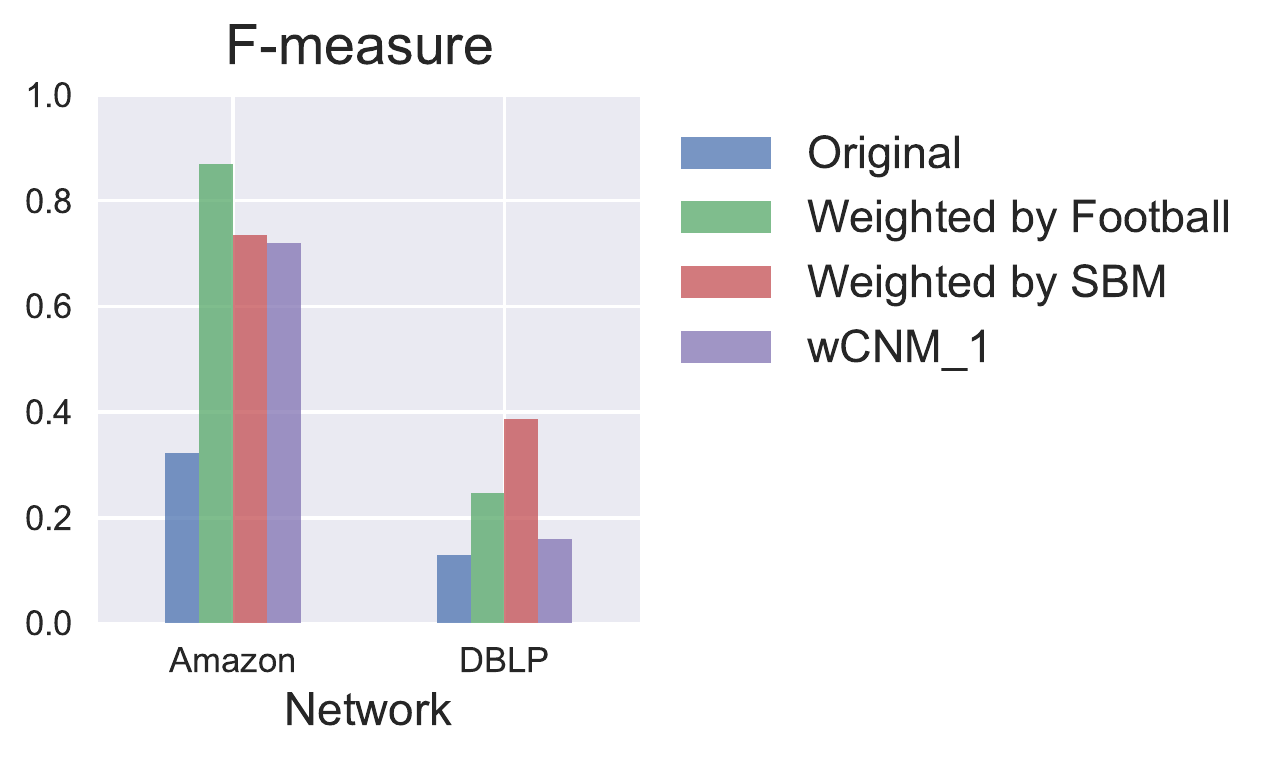}
        \caption{F-measure}
    \end{subfigure}
    \begin{subfigure}[b]{0.47\textwidth}
        \centering
        \includegraphics[width=\textwidth]{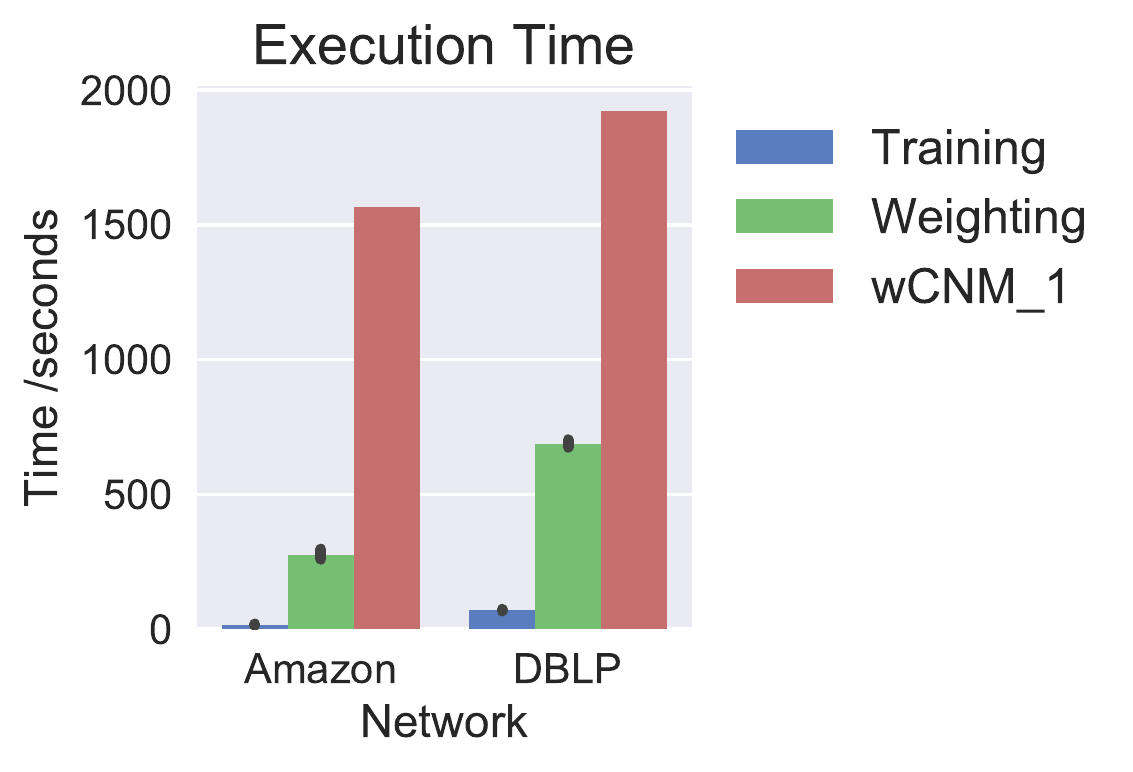}
        \caption{Efficiency}
    \end{subfigure}
    \caption{Performance improvement of community detection in Amazon and DBLP networks. (a-b) The number of communities detected by the modularity maximization in weighted and unweighted graphs in relation to the community size. Only communities containing more than 3 nodes are considered. (c) F-measure of the detected communities in the weighted graphs produced by our model with either Football network or artificial SBM networks used as the training data. (d) Time spent on the proposed training algorithm (i.e., Training), edge weighting (i.e., Weighting) and the wCNM\_1~\cite{berry2011tolerating} algorithm using a machine with 2.5 GHz Intel Core i5 CPU and 4GB memory.}
    \label{fig:comm_distribution}
\end{figure}

We evaluate the performance of our model on two large real networks: Amazon co-purchasing network and DBLP co-authorship network. The Amazon co-purchasing network~\cite{yang2015defining} consists of 334,863 products with two frequently co-purchased products linked by an undirected edge. Each collection of products from the same category forms one ground-truth community. The DBLP collaboration network~\cite{yang2015defining} is the co-authorship network where every node represents a researcher. Two researchers who published at least one paper together are linked. Following others, we assume that individual ground-truth community is defined by the publication venue, e.g., journal or conference. As seen in Figure~\ref{fig:comm_distribution}, this assumption is not correct. In case of large conferences and most of the journals, each researcher writes papers with only a small fraction of all authors publishing in a venue. Yet, each researcher is likely to write several papers with the same co-authors to a group of conferences and journals covering their research interests. Hence, we believe that real grand truth communities in the DBLP network are smaller than a set of authors for a single venue. In both networks, the top 5000 high-quality ground truth communities are provided and we compare them with the detected ones to compute the F-measure, as shown in Figure~\ref{fig:comm_distribution}.c.

The proposed weighting scheme is compared with the wCNM\_1 algorithm~\cite{berry2011tolerating} which computes the weight of an edge using all the triangles and 4-cycles containing it. In our experiments, the wCNM\_1 algorithm iterates only once over updates, because the results in~\cite{berry2011tolerating} show that additional iterations negligibly improve the final results.

The regression model which converts the original unweighted graphs to weighted ones is trained by sampling the dynamically constructed artificial SBM networks as described in Section~\ref{sec:conf}. In addition, we also test the performance of our model trained by the ground truth communities in the American college football network, as shown in Figure~\ref{fig:comm_distribution}. Perhaps surprisingly, the accuracy of the modularity maximization algorithm on the weighted graph when weights were based on SBM artificial network has improved for the Amazon network by almost 50\% as measured by F-score and even more for the DBLP network. 

The sizes of communities discovered in Amazon and DBLP networks containing more than $3$ nodes are plotted in Figure~\ref{fig:comm_distribution}a-b. In the weighted graph produced by our model for the Amazon network, the distribution of the sizes of the detected communities is close to the distribution of the sizes of ground truth communities for weights based on SBM artificial network but quite different for weights based on the Football network. Since the F-score was similar for those two cases, this result demonstrates the importance of inspecting the distribution of the community sizes. In case of the DBLP network, the improvement of F-score is significant for the weights based on the SBM network, but the distribution of the community sizes is different. We believe that these two results show that presumed ground truth communities in DBLP are not correct, and that smaller communities of researchers co-authoring papers across several venues are the right communities. These results show that our model successfully converts large networks to the weighted ones where the modularity maximization algorithms can perform better than they do on the original unweighted networks.

In general, these large networks can be processed in a few minutes as shown in Figure~\ref{fig:comm_distribution}d. The computation time is divided into two parts: (i) Training: the time spent to infer all the parameters of the regression model; (ii) Weighting: the time needed to compute the weights of every edge in the graph. Both steps include the I/O processing time of loading the network files from disk. The edge topological feature extraction (i.e., weighting) time increases as the number of edges grow, therefore processing of dense graphs can be more time-consuming. Unlike the weighting time, the training time does not change much with the size of the original network. This is because the size of the constructed artificial network is independent of the size of the original input graph. Last but not least, in our experiments, the edge topological feature extraction and edge weight evaluation use a single thread implementation. However, as problems that are easily parallelizable, they can be partitioned into many individual tasks to achieve a better performance.

\section{Conclusions} \label{sec:conclusions}
We have developed a novel regression model of assigning weights to edges to assist the discovery of community structures based on modularity maximization. Surprisingly, the results show the execution of Fast Greedy algorithm on a weighted graph improves $Q_{ds}$ value for the original unweighted graph. In other words, the weighted edges allows the algorithm to escape from the local maximum of $Q_{ds}$ in unweighted network and the solution found with the weighted edges has higher value of $Q_{ds}$ metric in the original unweighted network. Other community detection algorithms which are not based on the modularity maximization principle may also benefit from running on the weighted graph produced by our model rather than the original unweighted graph. Moreover, we introduce speedup improvements to accelerate the training of our regression model. Experimental results show that our approach significantly improves the quality of community detection in both real and synthetic networks.

\section*{Acknowledgements}
This work was supported in part by the Army Research Laboratory (ARL) under Cooperative Agreement Number W911NF-09-2-0053 (NS-CTA), by the Army Research Office (ARO), grant W911NF-16-1-0524, and by the Office of Naval Research (ONR) Grant No. N00014-15-1-2640. The views and conclusions contained in this document are those of the authors and should not be interpreted as representing the official policies either expressed or implied of the Army Research Laboratory or the U.S. Government.

%\bibliographystyle{apa}
%\bibliographystyle{plain}

%if Els journal, we can submit the PDF file.
\bibliography{mybibfile}
%else if arXiv, then we should use .bbl file instead of .bib file
%\printbibliography
%\input{output.bbl}

\newpage
\appendix
\section{Proof of Theorems 1, 2 and 3} \label{appendix:1}
\noindent \textbf{Theorem 1.} 
\begin{pf}
For any community $c_i$, 
\begin{equation}
W^{in}_{c_i} = \sum_{e\in E^{in}_{c_i} } w_e \geq \sum_{e\in E^{in}_{c_i} } 1 = |E_{c_i}^{in}|.
\end{equation}
Thus,
\begin{align}
    Q^{w}(G^w, C) & = \sum_{c_i \in C} \left[ \frac{W_{c_i}^{in} }{ W } - \left( \frac{W_{c_i}}{ 2 W}\right)^2 \right]\\
    & \geq \sum_{c_i \in C} \left[ \frac{|E_{c_i}^{in}| }{ |E| } - \left( \frac{d_{c_i}}{ 2 |E|}\right)^2 \right]\\
    & = Q(G, C).
\end{align}
\end{pf}
\textbf{Theorem 2.}
\begin{pf} 
Without loss of generality, consider two ground truth communities $c_i$ and $c_j$ \textit{enhanced} by the balanced edge weighting scheme. The change in modularity $Q^w$ upon joining these two communities is,
\begin{align} \label{eq:change}
\Delta Q^w_{c_i,c_j} &=  \frac{W_{c_i, c_j}}{W} - \frac{W_{c_i}W_{c_j}}{2W^2} = \frac{W_{c_i, c_j}}{|E|} - \frac{d_{c_i} d_{c_j}}{2|E|^2}\\
& \leq \frac{|E_{c_i, c_j}|}{|E|} - \frac{d_{c_i} d_{c_j}}{2|E|^2} = \Delta Q_{c_i,c_j}
\end{align}
\end{pf}

\noindent \textbf{Theorem 3.}
\begin{pf} 
Consider a community $c=c_i\cup c_j$ with $d_c \leq \sqrt{8|E|}$ \textit{enhanced} by the balanced edge weighting scheme, where $c_i$ and $c_j$ are two non-empty communities, then $\Delta Q_{c_i,c_j} \geq 0 $ by assumption that modularity reaches local maximum for partition $C^*$.

If $E_{c_i,c_j} = \emptyset$, then $W_{c_i, c_j} = 0$ and $ \Delta Q_{c_i,c_j} = - \frac{d_{c_i} d_{c_j}}{2|E|^2} \geq 0$. This leads to either $d_{c_i}=0$ or $d_{c_j}=0$ which causes contradiction. Otherwise, if $|E_{c_i,c_j}| \geq 1$, then $W_{c_i, c_j} \geq 1$ because the edge weighting scheme assigns weight $w_e \geq 1$ to any edge $e \in E_{c_i,c_j}$. Since $W_{c_i} + W_{c_j} = W_c = d_c$, we have $W_{c_i}W_{c_j} \leq (\frac{d_c}{2})^2$. When community $c$ splits into communities $c_i$ and $c_j$, the change in modularity $Q^w$ is,
\begin{align}
- \Delta Q^w_{c_i,c_j} &=  \frac{W_{c_i}W_{c_j}}{2W^2} - \frac{W_{c_i, c_j}}{W}\leq \frac{(\frac{d_c}{2})^2}{2W^2} - \frac{1}{W}\\
&= \frac{(d_c)^2 - 8 W }{8W^2} = \frac{d_c^2 - 8 |E| }{8|E|^2}\leq 0
\end{align}
Note that the last inequality holds because of the condition $d_c\leq \sqrt{8|E|}$.
\end{pf}

\end{document}